\documentclass{ws-procs975x65}            

\newcommand{\schw}{\big(1-\frac{2M}{r}\big)}
\newcommand{\Schw}{\Big(1-\frac{2M}{r}\Big)}
\newcommand{\eroot}{\sqrt{e^2-1+\frac{2M}{r}}}
\newcommand{\diag}{\operatorname{diag}}

\begin{document}

\title{Clarifying spatial distance measurement}

\author{Colin MacLaurin$^*$}

\address{School of Mathematics and Physics\\
University of Queensland, Australia\\
$^*$E-mail: colin.maclaurin@uqconnect.edu.au}

\begin{abstract}

We examine length measurement in curved spacetime, based on the 1+3-splitting of a local observer frame. This situates extended objects within spacetime, in terms of a given coordinate which serves as an external reference. The radar metric is shown to coincide with the spatial projector, but these only give meaningful results on the observer's $3$-space, where they reduce to the metric. Examples from Schwarzschild spacetime are given.

\end{abstract}


\section{Introduction and motivation}


Recall the textbook ``radial proper distance'' in Schwarzschild spacetime:
\begin{equation}
\label{eqn:dLSchw}
ds = \Schw^{-1/2}dr,
\end{equation}
which follows from setting $dt = d\theta = d\phi = 0$ in the line element in Schwarzschild coordinates. But what is the physical motivation for choosing a slice Schwarzschild $t = \textrm{const}$, rather than some other time coordinate? Special relativity stresses length is relative to the observer, so which observers measure Equation~\ref{eqn:dLSchw}, and what do others measure? The claim the $dt = 0$ slice is ``measurement by an observer at infinity'' is problematic, because there are many ways to extend a local frame, or to choose a simultaneity convention between distant frames.

Consider the same procedure repeated for Gullstrand-Painlev\'e coordinates:
\begin{equation}
\label{eqn:dLGP}
ds = dr.
\end{equation}
Painlev\'e hence concluded relativity is self-contradictory.\cite{painleve1921} Instead, as we shall see, these correspond to measurements by different observers. (Mathematically, note the differing expressions called ``$ds$'' are really restrictions to different subspaces.) We apply four complementary theoretical tools: suitably chosen coordinates, the spatial projector, the radar metric, and adapted frames.

\section{Well-suited coordinates}

An intuitive and pedagogical approach to length measurement is to provide coordinates suited to a given congruence of observers, if possible. Consider for example radial geodesic motion in Schwarzschild spacetime, with $4$-velocity field $\mathbf u$ parametrised by the Killing energy per mass $e := -\mathbf g(\mathbf u,\partial_t)$. A generalisation of Gullstrand-Painlev\'e coordinates has metric:\cite{gautreauhoffmann1978,martelpoisson2001,finch2015,bini+2012}
\begin{equation}
\label{eqn:GenGPmetric}
ds^2 = -\frac{1}{e^2}\Schw dT^2\mp\frac{2}{e^2}\eroot dT\,dr +\frac{1}{e^2}dr^2 + r^2(d\theta^2+\sin^2\theta\,d\phi^2),
\end{equation}
where $T \equiv T_e$ is the Einstein-synchronised proper time of the observers, which all share the same $e \in \mathbb R\backslash\{0\}$. For details see Ref.~\citenum{maclaurin2019domoschool}. Setting $dT = d\theta = d\phi = 0$ gives:
\begin{equation}
\label{eqn:dLeSchw}
dL = \frac{1}{e}dr,
\end{equation}
where we write $dL$ in place of $ds$ for the length element, and the sign choice is mere convention. The physical justification behind $dT = 0$ is that the $T = \textrm{const}$ hypersurface is orthogonal to the $4$-velocities, so coincides with their local $3$-spaces.

Note a static observer at $r = r_0$ is identical to an observer falling from rest at $r_0$, in the sense their velocities and hence local $3$-spaces coincide instantaneously. Both have $e = \sqrt{1-2M/r_0}$, so Equation~\ref{eqn:dLeSchw} reduces to the usual quantity (Equation~\ref{eqn:dLSchw}). Some textbooks set up a false dichotomy that $dr$ is not the distance but $(1-2M/r)^{-1/2}dr$ \emph{is}. Instead, for radial observers with $e = \pm 1$ the $r$-coordinate is precisely proper distance.

Equation~\ref{eqn:dLeSchw} is remarkably little known. Gautreau \& Hoffmann derived it, using a different parameter corresponding to the $0 < e < 1$ case.\cite{gautreauhoffmann1978} Taylor \& Wheeler present the $e = 1$ case clearly, which is the only textbook coverage apparently.\cite{taylorwheeler2000} Finch showed the $3$-volume inside the horizon is $1/e$ times its Euclidean value $\frac{4}{3}\pi(2M)^3$, for the $e > 0$ case. (Precedents include Lema\^itre, who pointed out $3$-space is Euclidean for $e = 1$, and mentioned measurement.\cite{lemaitre1932} Painlev\'e made the same observations, but mistakenly saw contradiction.\cite{painleve1921})

In general, consider a $4$-velocity field $\mathbf u$. Define a new coordinate $T$ by:
\begin{equation}
dT := -N^{-1}\mathbf u^\flat,
\end{equation}
where $\mathbf u^\flat$ is the $1$-form dual to the $4$-velocity, and $N$ is a lapse. $T$ exists locally iff the velocity gradient is vorticity-free, a consequence of Frobenius' theorem. Then the $T = \textrm{const}$ hypersurfaces are orthogonal to the congruence, since $dT(\boldsymbol\xi) = 0$ for any vector $\boldsymbol\xi$ orthogonal to $\mathbf u$.\footnote{If in addition the congruence is geodesic, we can set $N \equiv 1$, then $dT/d\tau = -\mathbf u^\flat(\mathbf u) = 1$, so $T$ measures proper time. This trick to derive $T$ was applied to relativity by Synge, and Lagrange's $3$-velocity potential in Newtonian mechanics is an antecedent} Now express the metric in coordinates including $T$, and set $dT = 0$.

\section{Spatial projector}

Given a $4$-velocity $\mathbf u$, the metric splits into parts parallel and orthogonal to $\mathbf u$ as $g_{\mu\nu} \equiv -u_\mu u_\nu + (g_{\mu\nu}+u_\mu u_\nu)$, assuming metric signature $-\textrm{+++}$. The latter term is the \emph{spatial projection tensor} $\mathbf P$, which extracts the spatial part of tensors via contraction. In particular, $P^\mu_\nu u^\nu = 0$, and $P^\mu_\nu\xi^\nu = \xi^\nu$ for any vector $\boldsymbol\xi$ in the $3$-space orthogonal to $\mathbf u$. Furthermore $P_{\mu\nu}\xi^\mu\xi^\nu = g_{\mu\nu}\xi^\mu\xi^\nu$ for such a $\boldsymbol\xi$, so $\mathbf P$ is also called the \emph{spatial metric}.

One may wonder if $dL'^2 := P_{\mu\nu}\xi^\mu\xi^\nu$ is meaningful as a length measurement for \emph{any} $\boldsymbol\xi$, not necessarily orthogonal to $\mathbf u$. For radial motion in Schwarzschild spacetime, the projector in Schwarzschild coordinates is
\begin{equation}
\label{eqn:projSchw}
P_{\mu\nu} =
\begin{pmatrix}
e^2-1+\frac{2M}{r} & e\schw^{-1}\eroot \\
e\schw^{-1}\eroot & \framebox{$e^2\schw^{-2}$} \\
\end{pmatrix}
\end{equation}
in the $t\text{-}r$ block, plus the usual $2$-sphere metric components $\diag(r^2,r^2\sin^2\theta)$ in the $\theta\text{-}\phi$ block. One might expect the radial direction to be the coordinate basis vector $\partial_r$, at least for $r > 2M$. Contracting with $\boldsymbol\xi = \partial_r$ picks out the $P_{rr}$ component (boxed), for a candidate spatial measurement:
\begin{equation}
\label{eqn:dLSchwcontracted}
dL =  e\Schw^{-1}dr.
\end{equation}

But consider the ``same'' contraction of tensors expressed in the generalised Gullstrand-Painlev\'e coordinates. The projector is:
\begin{equation}
\label{eqn:projGenGP}
P_{\mu\nu} = \begin{pmatrix}
\frac{1}{e^2}(e^2-1+\frac{2M}{r}) & \frac{1}{e^2}\eroot \\
\frac{1}{e^2}\eroot & \framebox{$\frac{1}{e^2}$} \\
\end{pmatrix},
\end{equation}
in the $T\text{-}r$ block, so the contraction with $\partial_r$ yields $dL = \frac{1}{e}dr$ as in Equation~\ref{eqn:dLeSchw}. But why the discrepancy with Equation~\ref{eqn:dLSchwcontracted}? While Equations~\ref{eqn:projSchw} and \ref{eqn:projGenGP} represent the same tensor, it turns out the coordinate vectors $\partial_r$ are distinct. By definition of coordinate basis, $^{(Schw)}\partial_r$ is orthogonal to $dt$, whereas $^{(GP)}\partial_r$ is orthogonal to $dT$. (In contrast, the vector $(dr)^\sharp$ depends only on $r$.) This potential confusion about coordinate vectors is rarely discussed explicitly.\cite{finch2015,maclaurin2019domoschool} In the present context, it shows a potential pitfall for measurement, and the superficial contradiction motivates deeper study.

In fact Equation~\ref{eqn:dLSchwcontracted} does have physical meaning: it is the measurement of a falling ruler as determined in the local static frame. By this, we mean the comparison of the ruler's length-contracted tick marks with the $r$-coordinate. The two results are related by the Lorentz factor $\gamma = |e|(1-2M/r)^{-1/2}$, since the frames are in standard configuration. This will be examined in future work. For now we conclude $\sqrt{\mathbf P_{\mathbf u}(\boldsymbol\xi,\boldsymbol\xi)}$ is \emph{not} a measurement in $\mathbf u$'s frame, if $\mathbf g(\boldsymbol\xi,\mathbf u) \ne 0$.

\section{Radar metric}

The sonar / radar method of distance measurement involves bouncing a signal off a distant object, and timing the return journey. In relativity a null signal is used, along with the proper time $\Delta\tau$ of the emitter, hence the one-way distance is $\Delta L := \Delta\tau/2$ (assuming an isotropic speed of light $c = 1$). While radar was promoted by Poincar\'e, Einstein, Milne, Bondi, and others, the following formula was derived by Landau \& Lifshitz:\cite{landaulifshitz1941}
\begin{equation}
\label{eqn:radarmetric}
\gamma_{ij} := g_{ij} - \frac{g_{0i}g_{0j}}{g_{00}},
\end{equation}
for $i, j = 1, 2, 3$, with infinitesimal length element $dL^2 = \gamma_{ij}dx^idx^j$. This assumes a particular coordinate system is provided, and that the radar instrument is \emph{comoving} in those coordinates.

To apply this to radial motion in Schwarzschild, we need comoving coordinates. One choice is a case of Lema\^itre-Tolman-Bondi coordinates, using $\rho_e$ given in Refs.~\citenum{gautreauhoffmann1978} and \citenum{maclaurin2019domoschool}, together with $T_e$ from above. The radar metric is:
\begin{equation}
\label{eqn:radarLem}
\gamma_{ij} = \begin{pmatrix}
\frac{1}{e^2}\Big(e^2-1+\frac{2M}{r}\Big) & 0 & 0 \\
0 & r^2 & 0 \\
0 & 0 & r^2\sin^2\theta
\end{pmatrix},
\end{equation}
with radial distance $dL = \frac{1}{e}\sqrt{e^2-1+2M/r}\,d\rho$. The interval ``$d\rho$'' is unfamiliar, hence reinterpret the radar metric as $4$-dimensional (which simply adds terms of $0$), and transform into other coordinates. It turns out Equation~\ref{eqn:radarLem} is identical to the spatial projector (Equations~\ref{eqn:projSchw} and \ref{eqn:projGenGP}), and the coordinate vector $\partial_\rho$ is parallel to both $^{(GP)}\partial_r$ and the radial ruler seen later (Equation~\ref{eqn:radialruler}).\footnote{Taking instead Schwarzschild $t$ with $\rho$, the radar metric gives Equation~\ref{eqn:dLSchwcontracted} upon contraction with the alternative $\partial_\rho'$. As before, coordinate vector directions can be misinterpreted}

In fact the spatial projector is the covariant generalisation of the radar metric. In any comoving coordinates, the observer has $4$-velocity $u^\mu = ((-g_{00})^{-1/2},0,0,0)$, assuming $\partial_0$ is future-pointing. Then $P_{\mu\nu}$ reduces to $\gamma_{\mu\nu}$, taking the latter as $4$-dimensional. Landau \& Lifshitz interpret $3$-space as spanned by the coordinate vectors $\partial_i$. However these are not necessarily orthogonal to $\partial_0$ and the $4$-velocity. Equation~\ref{eqn:radarmetric} gives incorrect results for directions not orthogonal to $\mathbf u$, as discussed for the spatial projector. When restricted appropriately, both radar and the projector are simply $g_{\mu\nu}$.


\section{Adapted frames}

Our final technical tool is local reference frames. The observer 4-velocity splits the local tangent space into ``time'' $\mathbf u$, and 3-dimensional ``space'' orthogonal to $\mathbf u$. For our purposes a single spatial vector $\boldsymbol\xi$ is often sufficient. While it is well known that measurements relate to expressing tensors in an observer's frame, strangely this is not applied to spatial distance --- with rare exceptions.\cite{klauber2004} Likewise Rindler writes, ``rigid scales [rulers] are of ill repute in relativity'', but there is nothing wrong with `` `resilient' scales'', within limits on acceleration and tidal forces.\cite{rindler1977}

For radial motion in Schwarzschild, the obvious choice of radial vector is that orthogonal to $\mathbf u$, $d\theta$ and $d\phi$:
\begin{equation}
\label{eqn:radialruler}
\xi^\mu := \Big(-\Schw^{-1}\eroot,e,0,0\Big).
\end{equation}
Since the $r$-component is $e$ and the vector has unit length, this means a coordinate interval $\Delta r = e$ corresponds to a proper length $\Delta L = 1$. Hence $dL = \frac{1}{e}dr$ as before.\footnote{Our ``rulers'' are vectors in a single tangent space, which approximate short rulers over spacetime} In general, suppose a unit spatial vector $\boldsymbol\xi$, and a scalar $\Phi$ (for instance a coordinate) are provided. The change in $\Phi$ over the extent of the ruler is $\Delta\Phi = d\Phi(\boldsymbol\xi)$, hence:
\begin{equation}
\label{eqn:dPhimeasurement}
dL = \frac{1}{d\Phi(\boldsymbol\xi)}d\Phi = \frac{1}{(\xi^\Phi)}d\Phi,
\end{equation}
where the latter expression applies if $\Phi$ is taken as a coordinate. We require $d\Phi(\boldsymbol\xi) \ne 0$, meaning $\Phi$ is not constant along the ruler direction, so can demarcate length. In any spacetime, given a coordinate expression of a tetrad, one can simply read off a coordinate length interval by inverting the relevant component.

The ``best'' ruler direction to choose is typically not obvious. Given $\Phi$ and $\mathbf u$, one candidate is a certain maximal direction as follows. Recall the gradient vector $(d\Phi)^\sharp$ is the direction of steepest increase of $\Phi$ per unit length. However this is generally not \emph{purely} spatial, according to $\mathbf u$. Hence, restrict $d\Phi$ to $\mathbf u$'s $3$-space, then take the vector dual. This vector shows the fastest increase of $\Phi$ along any possible ruler of $\mathbf u$. The measurement turns out to be $(g^{-1}(d\Phi,d\Phi)+(d\Phi(\mathbf u))^2)^{-1/2}d\Phi$, or:
\begin{equation}
dL_{\textrm{max-}\Phi} = \frac{1}{\sqrt{g^{\Phi\Phi}+(u^\Phi)^2}}d\Phi,
\end{equation}
if $\Phi$ is a coordinate. Alternatively, given $\Phi$, we can ask which observers $\mathbf u$ make extremal measurements. If a $\Phi = \textrm{const}$ slice is spatial ($g^{\Phi\Phi} > 0$), then observers with $u^\Phi \equiv d\Phi/d\tau = 0$ are possible, these measure $(g^{\Phi\Phi})^{-1/2}d\Phi$ in the direction $(d\Phi)^\sharp$. No ruler orientation for observers with $u^\Phi \ne 0$ can achieve this.

Consider a Schwarzschild observer parametrised by both $e$ and the Killing angular momentum per mass $\ell$. Its maximum possible $r$ and $\phi$ measurements are:
\begin{equation}
dL_{r\textrm{-max}} = \Big(e^2-\Schw\frac{\ell^2}{r^2}\Big)^{-1/2}dr, \qquad dL_{\phi\textrm{-max}} = \frac{r\,d\phi}{\sqrt{1+\ell^2/r^2}},
\end{equation}
where motion is in the plane $\theta = \pi/2$. Incidentally, the corresponding ruler vectors are not orthogonal to one-another.
Note that while $r$ is known as the reduced circumference,\cite{droste1917} the Euclidean $2$-sphere measurement $r\,d\phi$ only holds for zero angular momentum $\ell = 0$, under this natural ruler orientation.

Our orthonormal frame approach is trivially the zero-distance limit of Fermi coordinates, however very few exact Fermi coordinate expressions are known.\cite{bini+2005}

\section{Discussion}

There are many potential questions and objections. It is non-trivial to start from textbook material such as $\int ds$ or first-principles radar. Despite some excellent books,\cite{defelicebini2010,jantzen+2013} our results fill an independent niche.

Given two events, why not extremise the length of spatial geodesics between them? This is a $4$-dimensional approach, whereas ours uses a 1+3-dimensional splitting, measuring within the rest space of a given observer.

Is the result coordinate-dependent? The coordinate $\Phi$ serves as an extrinsic standard, but the length element is simply the ruler $\boldsymbol\xi^\flat$, which is also the metric interval restricted to a $1$-dimensional subspace. 

In Schwarzschild, $r$ is timelike inside the horizon, so how can it describe spatial measurement? Indeed the normal vector $(dr)^\sharp$ is timelike, but the measurement direction $\boldsymbol\xi$ is spatial.


\bibliographystyle{ws-procs975x65}
\bibliography{biblio}

\end{document}